\documentclass[pra,amsmath,amssymb, twocolumn, showpacs]{revtex4}

\usepackage{dcolumn}

\usepackage{bm}

\usepackage[dvips]{graphicx}

\usepackage{epsfig}

\begin{document}

\title{Geometric phases for $N$-level systems through unitary integration}

\author{D.\ B.\ Uskov and A.\  R.\  P. Rau$^{*}$}
\affiliation{Department of Physics and Astronomy, Louisiana State University,
Baton Rouge, Louisiana 70803-4001}


\begin{abstract}
 
Geometric phases are important in quantum physics and are now central to fault tolerant quantum computation. For spin-1/2, the Bloch sphere $S^2$, together with a U(1) phase, provides a complete SU(2) description. We generalize to $N$-level systems and SU($N$) in terms of a $2(N-1)$-dimensional base space and reduction to a $(N-1)$-level problem, paralleling closely the two-dimensional case.  This iteratively solves the time evolution of an $N$-level system and gives $(N-1)$ geometric phases explicitly. A complete analytical construction of a $S^4$ Bloch-like sphere for two qubits is given for the Spin(5) or SO(5) subgroup of SU(4). 
\end{abstract}


\maketitle

\date{\today}
Coupled quantal systems with $N$ states, and their time evolution are of interest in broad areas of physics. Laser coupling between three or more atomic or molecular states, oscillations between the three flavors of neutrinos, and logic gates in quantum computation (or cryptography and teleportation) are but a few examples of the widespread interest in few-level systems. In a series of papers \cite{our set} with specific applications to three- and four-level problems of quantum optics and quantum information, we have explored a semi-analytic technique of ``unitary integration" \cite{Wei-Norman} for the time dependent operator equations involved. For Hermitian Hamiltonians, unitarity is preserved at each step, no matrix inversions are involved, and often the problem reduces to solving a single or a small set of Riccati equations for classical functions. The method also extends readily to non-unitary evolution with dissipation and decoherence \cite{Wendell, our set} that occurs in quantum computation. The evolution operator is written as a product of exponentials, each exponent containing one of the operators of the algebra and a multiplicative classical function of time, these functions chosen appropriately to solve the time-dependent  equations of interest. 

The increasing number of functions and coupled equations that they obey (for $m$ qubits, $N=2^m$) may have discouraged more applications of unitary integration. This Letter presents a solution by giving a systematic and compact scheme for general $N$, in a stepwise reduction that parallels the $N=2$ single spin case. We also connect to another area of research, the study of phases that are central to the fields of quantum information and quantum computing. ``Geometric" phases that depend only on global features \cite{Berry} may be especially important for quantum computing, possibly providing high fidelity and fault-tolerant operation \cite{fault-tolerant}. Berry's initial discussion in terms of an adiabatic evolution and a single phase described by U(1) symmetry have been generalized extensively \cite{Wilczek-Zee}. Mathematicians describe it as a holonomy on a fiber bundle \cite{fiberbundle}. 

What has been missing, however, is a simple description of the parameters and, specifically, phases for a $N$-level problem of the sort available for two levels. For the group SU(2) of a single spin, or qubit, two coordinates of a ``Bloch" sphere $S^2$ \cite{Blochsphere}, together with a U(1) phase, provide a complete description. In the language of differential geometry \cite{diffgeom}, SU(2) is thereby viewed as a fiber bundle of the base manifold $S^2$ and the U(1) fiber \cite{fiberbundle}. Generalizations to larger groups, especially SU($N$) groups that describe higher spins or multiple qubits, are desirable but are less straightforward \cite{Ben-Aryeh}. For instance, a two-qubit system for logic gates \cite{gates} has SU(4) symmetry.  Bloch vectors for $N=3, 4$, and general $N$ \cite{Kimura} have been given but not the phase aspects and explicit evaluation of the evolution operator that we now provide. While having much in common with more mathematical treatments in \cite{Ben-Aryeh, Giavarini}, our emphasis is on concrete constructions through nothing more complicated than matrix algebra. 

The phase of a state's wave function being only accessible in comparison with another reference state, we find more useful the ``operator-valued phase", which may be called the ``$\mathcal{U}$-phase", of the evolution operator $U(t)$ since it is referenced to the unit operator at some initial time. The continuous connection to the starting unit operator also means that there are no $2n\pi$ ambiguities in these ``non-modular" phases \cite{Ben-Aryeh, Bhandari}. Our method of ``unitary integration" \cite{our set, Wei-Norman} expresses $U$ as a product of exponentials, each with one of the $(N^2-1)$ generators in the exponent. Thus, for SU(2), $U$ is a product of three exponentials in the Pauli matrices, $\sigma_{+}, \sigma_{-}, \sigma_z $, the last of which provides the $\mathcal{U}$-phase which can be further divided into dynamical and geometrical parts. For SU($N$), we construct a similar product form of two factors involving nilpotent exponents for the base and a third factor that is block diagonal in $(N-n) \times (N-n)$ and $n \times n$ blocks for the fiber. 

The use of nilpotent operators is central to our derivation. The complex time-dependent coefficients multiplying each operator in an exponent obey coupled first-order, initial value differential equations. For SU(2), a single complex parameter, $z(t)$, that multiplies $\sigma_{+}$ (we define $\sigma_{\pm} \equiv \sigma_x \pm i \sigma_y$, differing by a factor of 2 from some others) can be solved through a single Riccati equation for a vector on the Bloch sphere. This $z$ then provides through straightforward integration the remaining phase along with its breakdown into geometrical and dynamical parts. For a spin-1/2, charged particle in a magnetic field, with $H=-\frac{1}{2} \vec{\sigma} \cdot \vec{B}(t)$, solution of the Schr\"{o}dinger equation, $i\dot{U}(t)=H(t)U(t), U(0)= \mathcal{I}$, can be written as (an overdot will denote differentiation with respect to time)

\begin{equation}
U(t) = e^{z(t) \sigma_{+}/2} e^{w^{*}(t) \sigma_{-}/2} e^{- i\mu(t) \sigma_z/2}.
\label{eqn1}
\end{equation}
The complex quantities $z$, $w$, and $\mu$ are classical functions of time, vanishing at $t=0$. Solving the Schr\"{o}dinger equation with this form amounts to solving a Riccati equation for $z$  \cite{our set}. Unitarity of $U$ requires: $w^{*}=-z^{*}/(1+|z|^2), \,\, e^{{\rm Im}\, \mu} =  (1+|z|^2)$. Thus, there are only three linearly independent quantities, the real and imaginary parts of $z$ and Re $\mu$, the last being determined by quadrature in terms of $z$. $U$ depends on three parameters but the density matrix on $z$ alone.  

A specific focus on the phase $\mu$ follows upon viewing the evolution operator $U$ in Eq.~(\ref{eqn1}) as $\tilde{U}_1\tilde{U}_2$, where $\tilde{U}_2$ is the last factor involving $\mu$ and is diagonal,
whereas $\tilde{U}_1$ is given by the first two factors and depends on $z$ alone.
With such a product form, the Schr\"{o}dinger equation, $i\dot{U}=HU$,  reduces to one for $\tilde{U}_2$ alone with an effective Hamiltonian,

\begin{equation}
i\dot{\tilde{U}}_2 = H_{\rm eff} \tilde{U}_2, \,\,H_{\rm eff}=\tilde{U}_1^{-1}H\tilde{U}_1 -i\tilde{U}_1^{-1}\dot{\tilde{U}}_1.
\label{eqn2}
\end{equation}
This gives the equation for $\mu$. While the factorization of $U$ into two parts is generally valid, the SU(2) example displays explicitly the $z$ and $S^2$ Bloch sphere part and the $\mu$ phase, respectively. Note that $\tilde{U}_1$ and $\tilde{U}_2$ are not individually unitary, and we use the symbol tilde to signify this. However, as will be relevant for later development, factoring $\tilde{U}_2$ into real and imaginary parts, and incorporating the former into $\tilde{U}_1$ as a right multiplier gives a decomposition of the full $U$ into two unitary factors, $U_1$ and $U_2$, the first dependent on $z$ alone and the second on Re $\mu$. These are, respectively, operations on the base manifold and fiber of SU(2) regarded as a fiber bundle [SU(2)/U(1)] $\times$ U(1).

By defining a vector $\vec{m}$ on the Bloch sphere,

\begin{eqnarray}
m_{+} & \equiv & m_1 +im_2 = -2z^{*}/(1+|z|^2), \nonumber \\
m_3 & = & (1-|z|^2)/(1+|z|^2),
\label{eqn3}
\end{eqnarray}
with unit length: $m_1^2 +m_2^2 + m_3^2 = m_{+}m_{-} +m_3^2 =1$, one verifies the Bloch equation of motion, $\dot{\vec{m}} = -2 \vec{B} \times \vec{m}$,
with a Coriolis-like appearance. The two terms $\tilde{U}_1^{-1}H\tilde{U}_1$ and $-i\tilde{U}_1^{-1}\dot{\tilde{U}}_1$ of the effective Hamiltonian in Eq.~(\ref{eqn2}) have equal and opposite $\sigma_x$ and $\sigma_y$ terms which thereby cancel. The effective Hamiltonian reduces to a one-dimensional one in $\sigma_z$ for the fiber, the two parts, dynamical and geometrical, adding to give the full phase $\mu$. The dynamical part has the expected structure of the coupling energy of a magnetic moment to the magnetic field, $-\frac{1}{2} \vec{m} \cdot \vec{B}$. At the end of this Letter, we cast a non-trivial problem of two spins into an analogous five-dimensional unit vector moving on the sphere $S^4$ according to a generalized linear Bloch equation.

The above analysis extends to more general unitary Lie groups $\mathcal{U}$, which may not necessarily be U($N$) or SU($N$). For a sub-group $\mathcal{B}$ of $\mathcal{U}$, $\mathcal{B} \subset \mathcal{U}$, the factorization as before,
$U(t) =U_1(t) \,U_2(t), U_2 \in \mathcal{B}, \,\,\, U_1, U \in \mathcal{U}$,
holds, along with Eq.~(\ref{eqn2}). The factors are not unique, since for any $b \in \mathcal{B}$, we can rewrite as $U=(U_1b) U_2^{'}$ with $U_2 ^{'} =b^{-1} U_2$ again an element of $\mathcal{B}$. Thus, $U_1$ is determined up to right-multiplication by an element $b$, $U_1b$ representing one element in the space of left cosets, $\mathcal{U}/\mathcal{B}$. Because $b$ is a single phase when the group $\mathcal{B}$ is U(1), this non-uniqueness is like a gauge degree of freedom. Later below, we will encounter $b$ as a full matrix operator.
When $\mathcal{U}$ is SU($N$), and sub-group $\mathcal{B}$ is SU($N-1$), the above serves to define the fiber bundle [SU($N$)/(SU($N-1$) $\times$ U(1))] $\times$ (SU($N-1$) $\times$ U(1)),
which is the $n=1$ case of a more general construction 
[SU($N$)/(SU($N-n$) $\times$ SU($n$))] $\times$ (SU($N-n$) $\times$ SU($n$)) with $n \leq N/2$. The base manifold within square brackets is the Grassmannian manifold \cite{Grassmann} $Gr_{c} (N,n)$, elements of which can be represented by projection operators onto the $n$-dimensional subspace of $\cal{C}^{N}$. 

Thus, consider the $N$-dimensional Hamiltonian ${\bf H}^{(N)}$:

\begin{equation}
{\bf H}^{(N)} =\left(
\begin{array}{cc}
\tilde{\bf H}^{(N-n)} & {\bf V} \\
{\bf V}^{\dagger} & \tilde{\bf H}^{(n)}
\end{array}
\right).
\label{eqn4}
\end{equation}
The diagonal blocks are square matrices while the off-diagonal ${\bf V}$ is $(N-n) \times n$ and ${\bf V}^{\dagger}$ is $n \times (N-n)$. These latter are taken as Hermitian adjoints and ${\bf H}^{(N)}$ as traceless for purposes of this Letter although much of our construction applies more generally. We write, analogously to Eq.~(\ref{eqn1}), ${\bf U}^{(N)}(t) = \tilde{U}_1 \tilde{U}_2$, with 

\begin{eqnarray}
\tilde{U}_1 & = & \left(
\begin{array}{cc}
{\bf I}^{(N-n)} & {\bf z}(t) \\
{\bf 0}^{\dagger} & {\bf I}^{(n)}
\end{array} \right) \left(
\begin{array}{cc}
{\bf I}^{(N-n)} & {\bf 0} \\
{\bf w}^{\dagger}(t) & {\bf I}^{(n)}
\end{array} \right), \nonumber \\
\tilde{U}_2 & = &  \left(
\begin{array}{cc}
\tilde{\bf U}^{(N-n)} (t) & {\bf 0} \\
{\bf 0}^{\dagger} & \tilde{\bf U}^{(n)} (t)
\end{array} \right),
\label{eqn5}
\end{eqnarray}
where ${\bf z}$ and ${\bf w}^{\dagger}$ are rectangular matrices of complex parameters. Note that $\tilde{U}_1$ is constructed from block-matrix generalizations of $\sigma_{\pm}$ in Eq.~(\ref{eqn1}) and has blocks of zero in the lower and upper off-diagonal blocks of its matrix factors. Further, ${\bf U}^{(N)}$ as a whole is unitary which leads to matrix renderings of the relations noted earlier between $z$ and $w$:

\begin{eqnarray}
{\bf z} & = & -\mbox {\boldmath $\gamma$}_1 {\bf w}, \nonumber \\
\mbox {\boldmath $\gamma$}_1 = \tilde{\bf U}^{(N-n)} \tilde{\bf U}^{(N-n)\dagger} & = & {\bf I}^{(N-n)} +{\bf z}{\bf z}^{\dagger}, \nonumber \\
(\mbox {\boldmath $ \gamma$}_2)^{-1} = \tilde{\bf U}^{(n)} \tilde{\bf U}^{(n)\dagger} & = & ({\bf I}^{(n)} +{\bf z}^{\dagger} {\bf z})^{-1}.
\label{eqn6}
\end{eqnarray}

The effective Hamiltonian in Eq.~(\ref{eqn2}) constructed from $\tilde{U}_1$ is block diagonal, the off-diagonal blocks of the construction defining the equation satisfied by ${\bf z}$,

\begin{equation}
i\dot{\bf z}=\tilde{\bf H}^{(N-n)} {\bf z} + {\bf V}-{\bf z}({\bf V}^{\dagger}{\bf z}+\tilde{\bf H}^{(n)}),
\label{eqn7}
\end{equation}
now a matrix Riccati equation \cite{Reid}, and the diagonal blocks giving

\begin{equation}
\tilde{H}_{\rm eff} =  \left(
\begin{array}{cc}
\tilde{\bf H}^{(N-n)} -{\bf z}{\bf V}^{\dagger} & {\bf 0} \\
{\bf 0}^{\dagger} & \tilde{\bf H}^{(n)}+{\bf V}^{\dagger}{\bf z}
\end{array} \right),
\label{eqn8}
\end{equation}
the effective Hamiltonian for the $\tilde{U}_2$ problem. The individual blocks are neither Hermitian nor traceless.

To convert Eq.~(\ref{eqn5}) into a unitary decomposition, that is, to make both factors $U_1$ and $U_2$ unitary, we construct

\begin{equation}
\tilde{U}_1^{\dagger} \tilde{U}_1 =  \left(
\begin{array}{cc}
\mbox {\boldmath $ \gamma$}_1^{-1} & {\bf 0} \\
{\bf 0}^{\dagger} & \mbox {\boldmath $ \gamma$}_2
\end{array} \right),
\label{eqn9}
\end{equation}
and choose its inverse square root as the ``gauge factor" $b$ that provides the requisite unitarization: $U_1 \! =\tilde{U}_1 b$. 
Correspondingly, $U_2=b^{-1}\tilde{U}_2$ is unitary as well. Also, using this $U_1$ in an equation analogous to Eq.~(\ref{eqn2}) but without tildes, we get the explicitly Hermitian $H_{\rm eff}$ counterpart of Eq.~(\ref{eqn8}). The upper diagonal block in such an $H_{\rm eff}$ is

\begin{equation}
\frac{i}{2} [\frac{d}{dt} (\mbox {\boldmath $ \gamma$}_1 ^{-1/2}), \mbox {\boldmath $ \gamma$}_1^{1/2}]\! + \frac{1}{2} \{\mbox {\boldmath $ \gamma$}_1^{-1/2} (\tilde{\bf H}^{(N-n)}\!\!-{\bf z}{\bf V}^{ \dagger}) \mbox{\boldmath $ \gamma$}_1^{1/2} \!\! +{\rm h.c.} \},
\label{eqn10}
\end{equation}
with h.c.\ the Hermitian conjugate of the previous expression. The lower diagonal block of $H_{\rm eff}$ is
 
\begin{equation}
\frac{i}{2} [\frac{d}{dt} (\mbox {\boldmath $ \gamma$}_2 ^{-1/2}), \mbox {\boldmath $ \gamma$}_2 ^{1/2}]  \!+\! \frac{1}{2} \{ \mbox {\boldmath $ \gamma$}_2 ^{-1/2} (\tilde{\bf H}^{(n)}\!\!+{\bf z}^{\dagger}{\bf V})\mbox{\boldmath $ \gamma$}_2 ^{1/2} \!\! +  {\rm h.c.} \}.   
\label{eqn11}
\end{equation}
These serve as the explicitly Hermitian effective Hamiltonians for the SU($N-n$) and SU($n$) problems. 

Together with the $2n(N-n)$ parameters in ${\bf z}$ that are obtained through solutions of Eq.~(\ref{eqn7}), the $(N^2-1)$ parameters of SU($N$) are thereby expressed in terms of the corresponding parameters in the smaller groups and one additional phase parameter between those two spaces of $N-n$ and $n$ dimensions. Our solution may be regarded as extending the Schwinger scheme \cite{Sakurai} which constructs higher spin-$j$ representations from those of spin-1/2. While that scheme, and the association of $(N=2j+1)$-level systems with spin $j$ is familiar, it deals with larger representations but of the same group SU(2) or SO(3). Our decomposition of the full SU($N$) into SU($N-n$) and SU($n$) through two nilpotent and one diagonal factor in Eq.~(\ref{eqn5}) exactly analogous to the similar Eq.~(\ref{eqn1}) for SU(2) extends the scheme to a complete solution of an arbitrary SU($N$) Hamiltonian in Eq.~(\ref{eqn4}).   

While the above constructs the fiber bundle for arbitrary $n$, we will now concentrate on the case $n=1$. The handling of the square root operators for the general case will be considered elsewhere but for $n=1$, when ${\bf V}$ and ${\bf z}$ are column vectors, $\mbox {\boldmath $\gamma$}_2$ in Eq.~(\ref{eqn6}) reduces to a number,
 
\begin{equation}
\gamma \equiv 1 + {\bf z}^{\dagger} {\bf z}.
\label{eqn12}
\end{equation}
The commutator in Eq.~(\ref{eqn11}) vanishes and the rest of this expression is easily evaluated to give 

\begin{equation}
i\dot{U}_{2,NN} = [H_{NN} +\frac{1}{2}({\bf V}^{\dagger}{\bf z}+{\bf z}^{\dagger}{\bf V})]U_{2,NN},
\label{eqn13}
\end{equation}
where the $1 \times 1$ lower corner of the evolution operator's $U_2$ matrix (that is, $U^{(1)}(t)$ in   Eq.~(\ref{eqn5}) without the tilde) is denoted by $U_{2,NN}$ and likewise the corner element of ${\bf H}^{(N)}$ in Eq.~(\ref{eqn4}) by $H_{NN}$. The equation is readily integrated and gives a pure phase. It complements the real factor $(1/\sqrt{\gamma})$ in the corresponding corner of $U_1$, these being analogous to the real and imaginary factors of $\mu$ for the SU(2) example. The geometrical contribution, $-iU_1^{-1}\dot{U}_1$, to this $\mathcal{U}$-phase gives in the square bracket in Eq.~(\ref{eqn13}), ${\gamma}^{-1}[{\bf z}^{\dagger}(\tilde{{\bf H}}^{(N-1)}-H_{NN}){\bf z}+({\bf z}^{\dagger}{\bf V}+{\bf V}^{\dagger}{\bf z})(1-\gamma/2)]$. A useful identity in these constructions is $\dot{\gamma}=i\gamma ({\bf V}^{\dagger}{\bf z}-{\bf z}^{\dagger}{\bf V})$.

In the effective Hamiltonian in Eq.~(\ref{eqn10}) for the remaining SU($N-1$), the square roots simplify in terms of $\gamma$,

\begin{eqnarray}
\mbox {\boldmath $ \gamma$}_1 ^{1/2} \!& = &\! {\bf I}^{(N-1)}\!+{\bf z}{\bf z}^{\dagger}/(\sqrt{\gamma}+1), \nonumber \\
\mbox {\boldmath $ \gamma$}_1 ^{-1/2}\! & = & \! {\bf I}^{(N-1)}\!-{\bf z}{\bf z}^{\dagger}/(\sqrt{\gamma}+\gamma),
\label{eqn14}
\end{eqnarray}
and we get

\begin{equation}
{\bf H}^{(N-1)}=\tilde{{\bf H}}^{(N-1)}-\frac{{\bf z}{\bf V}^{\dagger}+{\bf V}{\bf z}^{\dagger}}{\sqrt{\gamma}+1}-\frac{{\bf z}({\bf z}^{\dagger}{\bf V}+{\bf V}^{\dagger}{\bf z}){\bf z}^{\dagger}}{2(\sqrt{\gamma}+1)^2}.
\label{eqn15}
\end{equation}
Note that the term in square brackets in Eq.~(\ref{eqn13}) is, as expected, $-{\rm Tr}{\bf H}^{(N-1)}$, ${\bf H}^{(N)}$ as a whole being traceless and trace being preserved in our construction. ${\bf H}^{(N-1)}$ in Eq.~(\ref{eqn15}) is the starting point for the subsequent construction of ${\bf U}^{(N-1)}$ after subtracting its trace. Thus both the base manifold SU($N$)/(SU($N-1$) $\times$ U(1)), determined entirely by ${\bf z}$, and the fiber are constructed in a form that allows for iteration. A fiber bundle description can be developed in turn for this SU($N-1$), thus giving a hierarchical construction. Each step provides a U(1) $\mathcal{U}$-phase. Turning to a geometrical picture, in the case of $N=2, n=1$, the Bloch sphere $S^2$ described by $\vec{m}$ in Eq.~(\ref{eqn3}) is an inverse stereographic projection of the one-dimensional Riemann plane given by the complex number $z$. Similarly, now for general $N$, and $n=1$, an inverse stereographic projection of the $(N-1)$-dimensional complex Riemann plane of ${\bf z}$ in Eq.~(\ref{eqn7}) gives the higher dimensional generalization of the Bloch sphere. The vector equation may not always reduce to a Bloch-like linear equation for a unit vector on $S^N$ but we now turn to an example for $N=4, n=2$ when it does. 

Consider two spins or qubits with a Hamiltonian of Spin(5) or SO(5) symmetry. There are many such within the full SU(4) dynamical symmetry of two spins but a specific representation as a concrete example is $H(t)=F_{21}\sigma^{(2)}_z-F_{31}\sigma^{(2)}_y+F_{32}\sigma^{(2)}_x-F_{4i}\sigma^{(1)}_z\sigma^{(2)}_i+F_{5i}\sigma^{(1)}_x\sigma^{(2)}_i-F_{54}\sigma^{(1)}_y$, where the ten arbitrarily time-dependent coefficients $F_{\mu \nu}(t)$ form a $5 \times 5$ antisymmetric real matrix. (We will use $\mu,\nu=1-5$ and $i,j,k=1-3$ and summation over repeated indices.) This form describes several quantum optics and multiphoton problems of four levels driven by time-dependent electric fields. It has more general couplings between the four levels than has been considered extensively in coherent population transfer and other phenomena in a variety of molecular and solid state systems \cite{Delgado}. Only numerical solutions have so far been available but we can now provide a complete analytical solution. With $n=2$, such a Hamiltonian in Eq.~(\ref{eqn4}) has $2 \times 2$ diagonal blocks, ${\bf H}^{(1,2)}=(\mp F_{4k} -\frac{1}{2} \epsilon_{ijk} F_{ij}) \sigma_k$, and off-diagonal ${\bf V}=iF_{54} {\bf I}^{(2)}+F_{5i} \sigma_i$. The matrix Riccati equation in Eq.~(\ref{eqn7}) can be solved in terms of four real $ z_{\mu}=z_4,z_i$, with ${\bf z}=z_4 {\bf I}^{(2)} -iz_i \sigma_i$, obeying 

\begin{equation}
\dot{z}_{\mu}=F_{5\mu}(1-z_{\nu}z_{\nu})+2F_{\mu \nu}z_{\nu}+2F_{5\nu}z_{\nu}z_{\mu}.
\label{eqn16}
\end{equation}
${\bf V}$ and $ {\bf z}$ can also be rendered in terms of quaternions $(1,-i\sigma_i)$. $\mbox {\boldmath $ \gamma $}_1$ and $\mbox {\boldmath $ \gamma $}_2$ in Eq.~(\ref{eqn6}) become equal and proportional to a unit matrix, $(1+z_{\mu}z_{\mu}){\bf I}^{(2)}$, and the evaluation of Eq.~(\ref{eqn10}) and Eq.~(\ref{eqn11}) gives straightforwardly ${\bf H}_{\rm eff}^{(1,2)}={\bf H}^{(1,2)}-\epsilon_{ijk} z_i F_{5j} \sigma_k \mp F_{5j} z_4\sigma_j \pm F_{54 }z_i \sigma_i$. Remarkably, in terms of a five-dimensional unit vector $m$,

\begin{equation}
m_{\mu}=\frac{-2z_{\mu}}{(1+z_{\nu}z_{\nu})}, m_5=\frac{(1-z_{\nu}z_{\nu})}{(1+z_{\nu}z_{\nu})}, \mu,\nu=1-4, 
\label{eqn17}
\end{equation}
the nonlinear Eq.~(\ref{eqn16}) in $z$ becomes a simple, linear Bloch-like $\dot{m}_{\mu}=2F_{\mu\nu}m_{\nu}, \mu,\nu=1-5$. As in the single spin case, this represents an inverse stereographic projection, now from the four-dimensional plane $z \in R^4$ to the four-sphere $S^4$, and thus provides a higher-dimensional polarization vector for describing such two spin problems. Further details and application to specific problems in quantum information and nuclear magnetic resonance, as well as other $n=1$ and $n=2$ applications of our construction, will be given elsewhere. 

This work has been supported by the National Science Foundation Grant 0243473 and by a Roy P. Daniels Professorship.

\end{document}